\documentclass[reprint,superscriptaddress,amsmath,amssymb,prl,floatfix,footinbib]{revtex4-1}

\usepackage[pdftex]{graphicx}
\usepackage{bm}
\usepackage{color}
\usepackage{times}
\usepackage{amsmath}
\usepackage{amssymb}
\usepackage{mathtools}
\usepackage{soul}

\newcommand\underrel[3][]{\mathrel{\mathop{#3}\limits_{%
      \ifx c#1\relax\mathclap{#2}\else#2\fi}}}

\begin{document}

\title{Optimized Diffusion of Run-and-Tumble Particles in Crowded Environments}

% Use letters for affiliations, numbers to show equal authorship (if applicable) and to indicate the corresponding author
\author{Thibault Bertrand}
\email[Electronic address: ]{thibault.bertrand@upmc.fr}
\affiliation{Laboratoire Jean Perrin, UMR 8237 CNRS, Sorbonne Universit\'{e}, 75005 Paris, France}
\author{Yongfeng Zhao}
\affiliation{Laboratoire Mati\`{e}re  et Syst\`{e}mes Complexes, UMR 7057 CNRS, Universit\'{e} Paris Diderot, 75205 Paris, France}
\author{Olivier B\'{e}nichou}
\affiliation{Laboratoire de Physique Th\'{e}orique de la Mati\`{e}re Condens\'{e}e, UMR 7600 CNRS, Sorbonne Universit\'{e}, 75005 Paris, France}
\author{Julien Tailleur}
\affiliation{Laboratoire Mati\`{e}re  et Syst\`{e}mes Complexes, UMR 7057 CNRS, Universit\'{e} Paris Diderot, 75205 Paris, France}
\author{Rapha\"{e}l Voituriez}
\email[Electronic address: ]{voiturie@lptmc.jussieu.fr}
\affiliation{Laboratoire Jean Perrin, UMR 8237 CNRS, Sorbonne Universit\'{e}, 75005 Paris, France}
\affiliation{Laboratoire de Physique Th\'{e}orique de la Mati\`{e}re Condens\'{e}e, UMR 7600 CNRS, Sorbonne Universit\'{e}, 75005 Paris, France}

\date{\today}

% Single-column figure: width=8.6cm
% Two-column figure: width=17.2cm

\begin{abstract}
We study the transport of self-propelled particles in dynamic complex
environments. To obtain exact results, we introduce a model of
run-and-tumble particles (RTPs) moving in discrete time on a
$d$-dimensional cubic lattice in the presence of diffusing hard-core obstacles. We derive an explicit expression for the diffusivity of the
RTP, which is exact in the limit of low density of fixed obstacles. To
do so, we introduce a generalization of Kac's theorem on the mean
return times of Markov processes, which we expect to be relevant for a
large class of lattice gas problems.  Our results show the diffusivity
of RTPs to be nonmonotonic in the tumbling probability for low
enough obstacle mobility. These results prove the potential for
optimization of the transport of RTPs in crowded and disordered
environments with applications to motile artificial and biological
systems.
\end{abstract}

\maketitle

%%%%%%%%%%%%%%
% Introduction
%%%%%%%%%%%%%%

Run-and-tumble particles (RTPs) are a prototypical model of
self-propelled particles (SPPs) at the colloidal scale, which belongs
to the broader class of active matter systems
\citep{ramaswamy-arcmp-2010,bechinger-rmp-2016}. In its simplest form,
RTP trajectories consist of a sequence of randomly oriented
"runs"---periods of persistent motion in a straight line at a constant
speed---interrupted by instantaneous changes of direction, called
"tumbles", occurring at random with a constant rate. This canonical
model has played a pivotal role in the theoretical description of
self-propelled biological entities such as bacteria
\citep{schnitzer-pre-1993,berg-book-2004,dileonardo-pnas-2010,saragosti-pnas-2011,schwarz-linek-pnas-2012},
algae \citep{polin-science-2009}, eukaryotic cells
\cite{heuze-immunolrev-2013}, or larger scale animals
\cite{benichou-rmp-2011}.

Whereas systems in thermal equilibrium display a time-reversal symmetry,
this invariance is generically lost in active matter at the
microscopic scale because of the continuous consumption of
energy. Nevertheless, at a constant speed and tumbling rate, an isolated
RTP performs a random walk with diffusive scaling at large enough
time- and length scales which cannot be qualitatively distinguished from the
equilibrium dynamics of Brownian colloids. Hence, it is only through
their interactions with either other particles or the environment that
RTPs display non-equilibrium features. Interactions between SPPs can
indeed have spectacular consequences which have attracted a growing
interest over the last decade. For instance, dense active suspensions can display
large-scale collective motion in settings where long-range order would
be forbidden for equilibrium
systems~\cite{vicsek-prl-1995,deseigne-prl-2010,bricard-nature-2013,solon-prl-2015}. Another
non-thermal collective effect is the propensity of active particles to
cluster~\cite{theurkauff-prl-2012,buttinoni-prl-2013,palacci-science-2013}
or undergo
phase-separation~\cite{tailleur-prl-2008,fily-prl-2012,redner-prl-2013}
in the presence of purely repulsive interactions.

The interplay between active particles and their environment has also
attracted a lot of interest~\cite{bechinger-rmp-2016}. Most motile
biological systems such as bacteria or dendritic cells navigate
disordered and complex natural environments such as soils, soft gels
({\it e.g.} mucus or agar) or tissues. Recent simulations have explored the dynamics of active particles in the presence of quenched disorder as well as active baths \citep{pince-natcomm-2016,volpe-pnas-2017,zeitz-epje-2017,sandor-pre-2017,reichhardt-pre-2015}. In confined geometries, active particles
accumulate at the boundaries, at odds with the equilibrium Boltzmann
distribution. This has been observed for a variety of systems from
spherical and elongated particles in linear channels to bacteria in
spherical
cavities~\citep{berke-prl-2008,wensink-pre-2008,elgeti-epl-2009,elgeti-epl-2013,vladescu-prl-2014,bricard-natcomm-2015}. Such
non trivial interactions with obstacles can lead to effective trapping and thus have important consequences in
the dynamics of SPPs in disordered environments. Indeed, active
particles in the presence of static obstacles can display subdiffusive
dynamics~\citep{chepizhko-prl-2013b}. Trapping has been observed both
in models~\citep{tailleur-epl-2009,kaiser-prl-2012,kaiser-pre-2013}
and in experiments of biological or synthetic microswimmers
\citep{galajda-jbacterio-2007,guidobaldi-pre-2014,restrepo-perez-labchip-2014}.
It was shown in numerical simulations of RTPs moving through arrays of obstacles that 
trapping can lead to the existence of an optimal activity level for drift through the system \citep{reichhardt-pre-2014,reichhardt-jpcm-2018}.
More recently, the presence of disordered obstacles was shown to destroy
the emergence of large scale correlations preventing
flocking and swarming \citep{morin-natphys-2017}. Despite these
various observations, the generic analysis of the dynamics of a single
SPP in disordered environments remains mostly unexplored, and in
particular analytical results are largely missing.

In this letter, we introduce a minimal model of discrete time RTP
moving on a $d$-dimensional cubic lattice in the presence of diffusing
hard-core obstacles of density $\rho$, which model a potentially
dynamic disordered environment. In particular, this generalizes to
RTPs questions that have attracted a lot of attention for passively
diffusing particles \citep{nakazato-ptp-1980} and externally driven
tracers~\citep{benichou-prl-2013,benichou-prl-2014,leitmann-prl-2013,cividini-pre-2017}. We
determine analytically numerous observables characterising the
dynamics : the mean free run time, defined as the mean time between
consecutive collisions of the RTP with obstacles, the mean trapping
time of the RTP by obstacles, and the large-scale diffusion
coefficient of the RTP. This calculation is exact for fixed obstacles
in the limit of low obstacle density $\rho \to 0$, and remains
uniformly accurate in the tumble rate for finite values of $\rho$ and
mobile obstacles. Our analysis reveals the existence of a maximum of
the diffusion coefficient of the RTP as a function of the tumbling
rate, for a low enough mobility of obstacles.  Our approach is based
on a generalization of a theorem due to Kac on mean return times of
Markov processes \cite{aldous-mono-2002}, which was already shown to
have important applications in physics
\cite{condamin-pre-2005,benichou-epl-2005}. We show here that it
implies the following exact result: For fixed obstacles the mean free
run time is given by $\left<\tau_r\right>=1/\rho$, and is {\it
  independent of the tumbling rate} of the RTP. In addition, in the
case of moving obstacles, this result still holds for a proper choice
of microscopic collision rules and is \textit{ independent of the
  diffusion coefficient of the obstacles}, provided that the time step
of the obstacle dynamics is larger than that of the RTP. This
strikingly simple result has the potential to find a variety of
applications in general lattice gas problems.

\begin{figure}[h]
\centering
\includegraphics[width=0.4\textwidth]{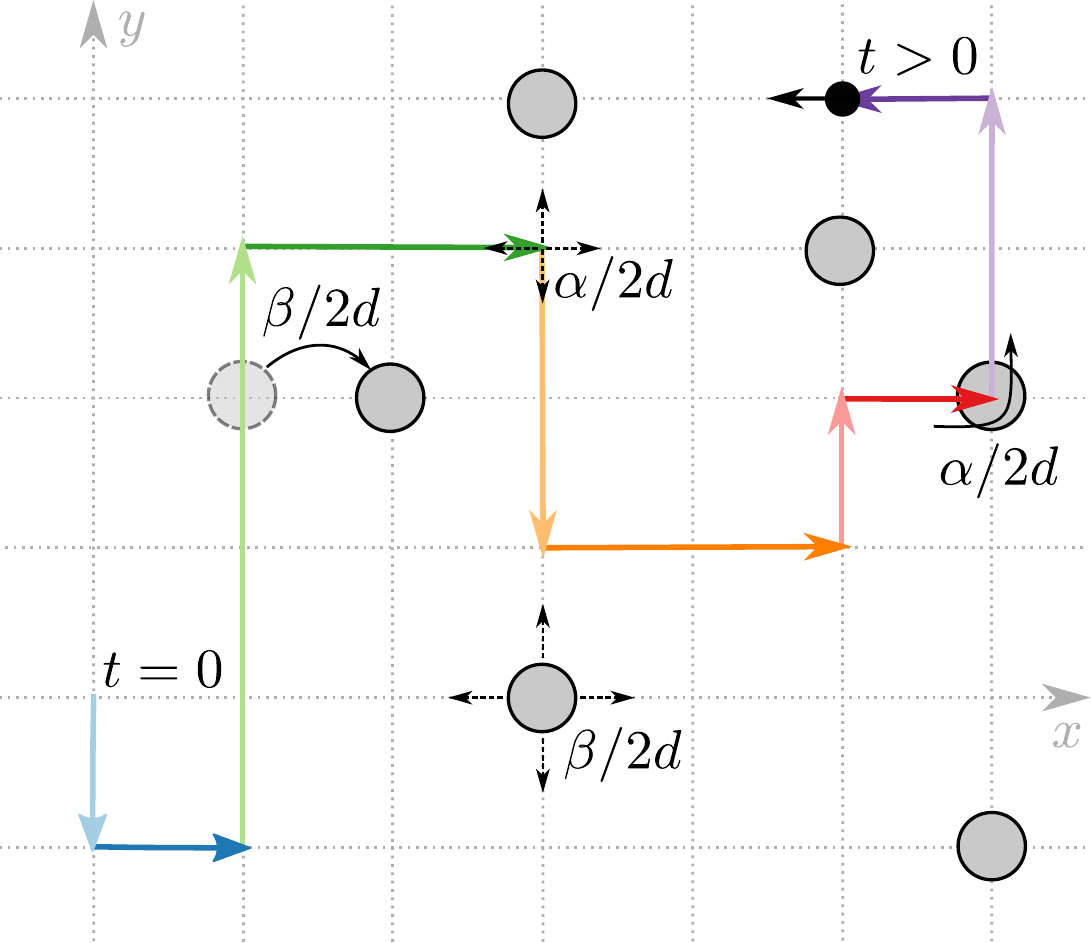}
\caption{{\it Example trajectory of a run-and-tumble particle} ({\it black}) on a 2D lattice among uniformly distributed obstacles ({\it gray}) at density $\rho$. At $t=0$, the RTP starts from the origin and it moves along the direction of its polarity ({\it black arrow}) in a sequence of linear runs (pictured here in different colors) punctuated by encounters with obstacles and tumbles. The obstacles have a probability $\beta/2d$ to move in a given direction  and the RTP has a probability $\alpha/2d$ to flip its polarity along a particular direction, potentially untrapping the RTP.}
\label{fig:trajectory}
\end{figure}

%%%%%%%%%%%
% Model definition   % 
%%%%%%%%%%%
{\it Model and definitions ---} We consider a discrete time RTP on an infinite lattice in $d$ dimensions surrounded by obstacles uniformly distributed with a density $\rho$ as shown in Fig. \ref{fig:trajectory}. The RTP, of position $\bm{r}(t)$, is polarized in a given direction and, in absence of interaction with obstacles,  makes one lattice step per unit time  in this direction (run) until its polarity is reset randomly amongst the $2d$ possible directions on the lattice (tumble). We  consider that these tumbling events happen at each time step with probability $\alpha$, independently of the presence of obstacles. 
We assume that  the obstacles perform symmetric nearest neighbour random walks,  with probability $\beta  <1$ to jump at each time step. We consider obstacles interacting via hard-core repulsion; i.e. each lattice site can contain at most one obstacle. The RTP is assumed to have hard-core interactions with obstacles; for the sake of simplicity, its size is, however, supposed negligible in front of the size of the obstacles. The RTP can therefore jump on a lattice site occupied by an obstacle, but cannot cross it. This assumption, illustrated in {\it Supplemental Material (SM)}, renders the analytical calculations more tractable but does not fundamentally change the phenomenology, as shown in {\it  SM}, where we analyzed the dynamics not allowing jumps of the  RTP on occupied sites. We therefore consider the following interaction rule: When the RTP steps on a lattice site occupied by an obstacle, it cannot proceed and effectively gets trapped; the RTP is released by either (i) a tumble leading to a change of polarity  or (ii) a step made by the obstacle in any direction.  Our goal is to compute analytically the diffusion coefficient $\mathcal{D} = \lim_{t\to\infty} \left< r^2(t)\right> / (2dt)$ as a function of the tumbling probability $\alpha$, the jump probability of the obstacles $\beta$ and the obstacle density $\rho$, where $\left< \ldots \right>$ represents an average over the trajectories. The dynamics starts  with a Poisson distribution of obstacles and we consider  timescales much larger than $1/\alpha$ and $1/\rho$, so that a stationary state is reached.

First, we decompose the trajectory of the RTP in a sequence of linear runs $\bm{a}_i$, punctuated by either encounters with obstacles, or tumbles in any direction:
\begin{equation}
\bm{r}(t) = \sum_{i=1}^{n(t)} \bm{a}_i
\label{eq:decomposition1}
\end{equation}
with the random variable $n(t)$ being the number of linear runs composing the trajectory up to a given time $t$. Hence, the RTP trajectory is a random sum of  random variables and the following exact asymptotics can be obtained by generalizing Wald's identity (see {\it SM}):
\begin{equation}
\left< r^2(t) \right> \underrel{t \to \infty}{\sim} \left< n(t) \right>\left< \bm{a}_i^2 \right> + \sum_{\substack{i,j=1 \\ i \ne j} }^{\left< n(t) \right>} \left< \bm{a}_i\cdot\bm{a}_j\right> .
\label{eq:decomposition2}
\end{equation}

We now  determine explicitly all terms involved in  (\ref{eq:decomposition2}). It is useful to  decompose a trajectory in  successive phases of two types: (i)  {\it mobile} phases of random duration $\tau_r$,  when the particle is freely moving on the lattice without interacting with obstacles ; (ii)  {\it static} phases of random duration $\tau_s$ when the particle is trapped by an obstacle.  For a trajectory of length $t$, the average number of each of the mobile and static phases is given by $\mathcal{N}_s = t/(\left< \tau_s\right>+\left< \tau_r\right>)$. We therefore deduce the mean number of runs performed until time $t$:
\begin{equation}\label{nbar}
\left< n(t) \right> \underrel{t \to \infty}{\sim}  \frac{t-\mathcal{N}_s \left< \tau_s\right>}{\ell_p} =  \frac{\left< \tau_r\right>}{(\left< \tau_s\right>+\left< \tau_r\right>)\ell_p}t \equiv \bar n t
\end{equation}
where the first (persistence length $\ell_p \equiv \left< | \bm{a}_i | \right>$) and second moment of the run length are given by 
\begin{equation}\label{a2}
\ell_p = \frac{1}{1-(1-\rho)(1-\alpha)},\ \left< \bm{a}_i^2\right> = \frac{1+(1-\rho)(1-\alpha)}{[1-(1-\rho)(1-\alpha)]^2}.
\end{equation}

In the case of moving obstacles, a trapped particle can be released by two competing independent mechanisms: tumbling of the RTP or stepping of the obstacle. Thus, the mean trapping time reads in the general case: 
\begin{equation}
\left< \tau_s \right> = \frac{1}{1-(1-\alpha^*)(1-\beta^*)} -1
\end{equation}
with probabilities $\alpha^* = \alpha(2d-1)/2d$ and $\beta^* = \beta(2d-1)/2d$.

%%%%%%%%%%%%%%
% Generalized Kac's theorem
%%%%%%%%%%%%%%
{\it Mean run time: Generalized Kac's theorem ---} We now determine the mean free running time $ \left< \tau_r \right>$. In the case of fixed obstacles, it can be exactly defined as the mean return time to the set $\cal O$ of all obstacles.  Remarkably, this quantity can be determined exactly by adapting Kac's theorem \cite{aldous-mono-2002} (see {\it SM} for details). For that purpose we introduce the auxiliary process  $\bm{\tilde r}(t)$. It  is identical to  $\bm{r}(t)$ in mobile phases, but the durations of all its static phases are set to 1 : Upon each trapping event by  an obstacle the auxiliary process is released in the same direction as the original process $\bm{ r}(t)$ would be but  after a single time step. The mean running time is then identical for both processes  $\bm{r}(t)$ and $\bm{\tilde r}(t)$; for the process $\bm{\tilde r}(t)$, which has a uniform stationary distribution, the Kac theorem takes a simple form and yields $ \left< \tau_r \right>=1/P_{\rm stat}({\cal O})$, where $P_{\rm stat}({\cal O})$ is the stationary probability of $\cal O$ for the auxiliary process. We therefore find the simple expression
\begin{equation}
\left< \tau_r \right> = \frac{1}{P_{\mathrm{stat}}({\cal O})} = \frac{1}{\rho},
\label{eq:kac}
\end{equation}
which is strikingly independent of the tumbling probability $\alpha$
and echoes results obtained on continuous space and time processes in
confined domains
\cite{blanco-epl-2003,benichou-epl-2005}. Interestingly, this result
can be generalized to the case of moving obstacles. To this end, we
encode the dynamics of the full system of $N$ obstacles of positions
${\bf r}_i(t)\ (1\le i\le N)$ and the auxiliary process $\bm{\tilde r}(t)$ in a $d(N+1)$ tuple
 ${\bf x}(t)$. The process ${\bf x}(t)$ performs a symmetric
random walk on the hyper cubic lattice of dimension $d(N+1)$, which
is, however, not of nearest neighbour type because several particles can
move in a given time step. Defining $\cal T=\{ {\bf x}, {\exists}$ $i,
{\bf r}_i={\bf r} \} $ as the set of trapped configuration, $ \left<
\tau_r \right>$ can be defined as the mean return time of the process
${\bf x}(t)$ to the set $\cal T$ and as such verifies $ \left< \tau_r
\right>=1/P_{\rm stat}({\cal T})$ in virtue of Kac theorem. Here
$P_{\rm stat}({\cal T})$ can depend on the specific choice of
microscopic collision rules between the RTP and obstacles. However, it
can be generically written $P_{\rm stat}({\cal T})=C\rho$, where the
constant $C$ is of the order 1 and can be exactly set to 1 for a proper
choice of microscopic rule~\footnote{{\it i.e.}, the choice of the
  order in the sequence of particle tumbles and particle moves during
  a given time step}. Here, we retain our initial choice, more
relevant to real motile systems, and show in {\it SM} a
very good agreement between our predictions and the results of
numerical simulations. This shows finally that, up to a redefinition
of microscopic interaction rules, the general expression
(\ref{eq:kac}) still holds for moving particles, showing that the mean
free running time is universally set by the density of obstacles only,
independently of both the tumbling probability of the RTP and the
dynamics of obstacles parametrised by $\beta$. This
result, key to the derivation below, has potential applications
to many other lattice gas models.

\begin{figure}[th!]
\centering
\includegraphics[width=0.46\textwidth]{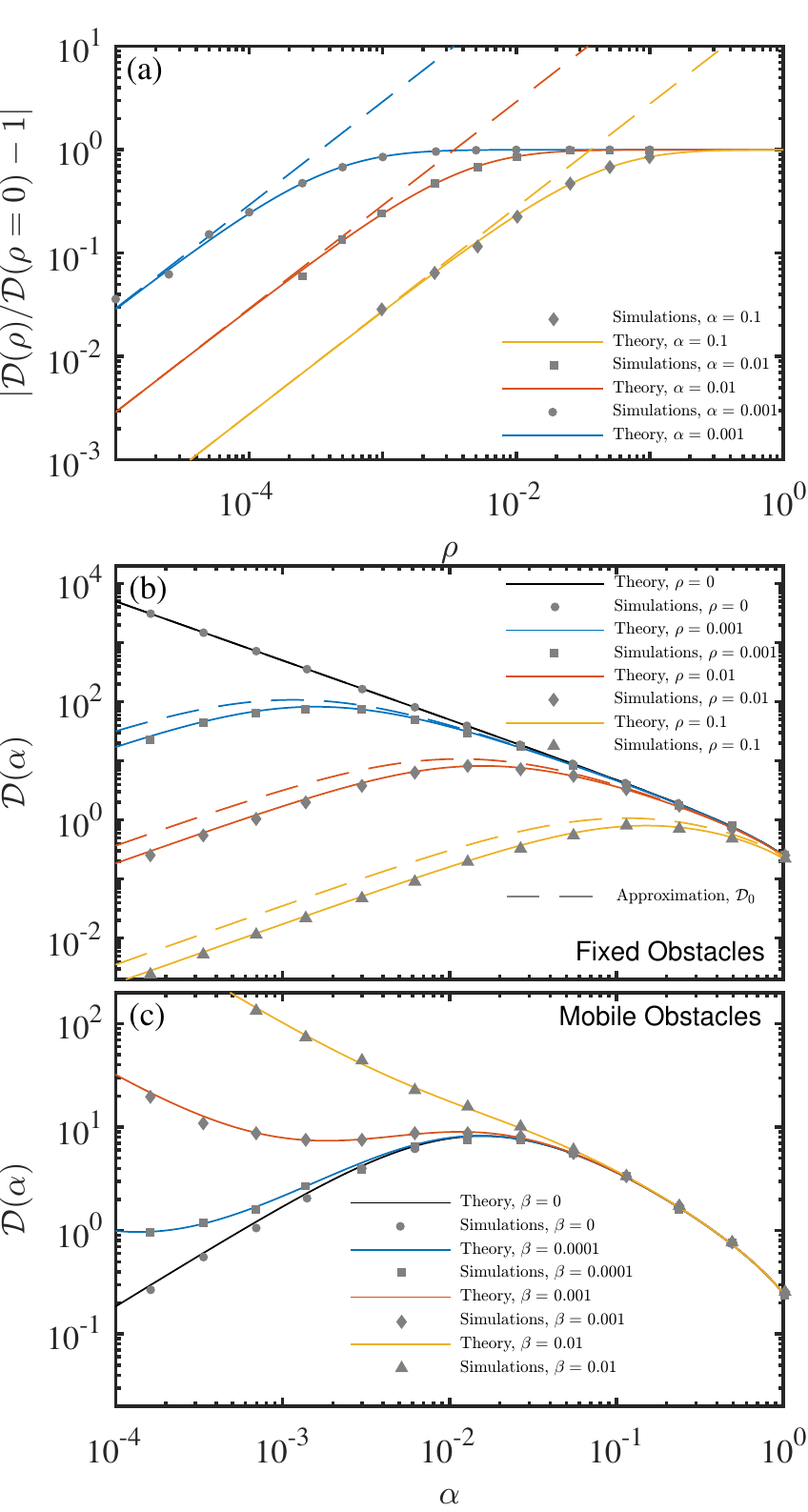}
\caption{{\it Diffusivity of a run-and-tumble particle} --- (a) Diffusion coefficients for fixed obstacles relative to the free diffusion coefficient $\mathcal{D} (\rho = 0) = (2-\alpha)/\alpha$ at various tumbling rates $\alpha$ as a function of the obstacles density $\rho$, dashed lines show the linear expansion in $\rho$. Diffusion coefficients for (b) fixed obstacles at various obstacle densities $\rho$ and (c) mobile obstacles at various obstacles mobilities $\beta$ and density $\rho = 0.01$ as a function of the tumbling probability $\alpha$. In both cases, the symbols are results of numerical simulations and the solid lines are given by Eq. \ref{eq:fulleffecivecoeffmobile}. In (b), dashed lines show $\mathcal{D}_0$, our approximation where correlations between runs are neglected.}
\label{fig:diffusion}
\end{figure}

%%%%%%%%%%%%%%
% Long time correlations  % 
%%%%%%%%%%%%%%
{\it Long time correlations ---} 
 As opposed to the classical RT dynamics in free space, obstacles  induce non trivial correlations $\left< \bm{a}_i\cdot\bm{a}_j\right>$ that remain to be determined to compute the diffusion coefficient of the RTP [see Eq.  (\ref{eq:decomposition2})]. A first approximation to the diffusion coefficient can be obtained by neglecting these correlations, yielding $\mathcal{D}_0  \equiv \bar{n} \left< \bm{a}_i^2\right> / (2d)$, where $\bar n$ and $\left< a_i^2\right>$ are determined exactly by Eqs. (\ref{nbar}) and (\ref{a2}). As shown in Fig. \ref{fig:diffusion}(b), $\mathcal{D}_0$ is already a qualitatively satisfactory approximation. Nevertheless, this approximation fails in the case of mobile obstacles and a more thorough treatment of the correlations is already necessary to obtain exact expressions even in the limit of low density of fixed obstacles. We thus treat exactly the case of fixed obstacles ($\beta=0$) to lowest order in $\rho$. The correlations can be qualitatively understood in the case of adjacent runs. If  $\bm{a}_i$ ends by a trapping event, then clearly $\left< \bm{a}_i\cdot\bm{a}_{i+1}\right><0$, because the obstacle forbids the run $\bm{a}_{i+1}$ to keep the direction of $\bm{a}_i$; alternatively if $\bm{a}_i$ ends by a tumble, one still finds  $\left< \bm{a}_i\cdot\bm{a}_{i+1}\right><0$, because particles are less likely to encounter obstacles upon retracing their steps.

More quantitatively,  an exact computation confirms this analysis and  yields (see {\it SM} for details):
\begin{equation}
\left< \bm{a}_i\cdot\bm{a}_{i+1}\right> = \frac{\alpha}{\alpha+\rho}\frac{ \mathcal{C}_+ -\mathcal{C}_-}{2d} - \frac{\rho}{\alpha+\rho}\frac{\mathcal{C}_-}{2d-1}
\label{eq:correlations}
\end{equation}								      
with the contribution to positive correlations $\mathcal{C}_+ = \ell_p^2$ and to negative correlations $\mathcal{C}_- = [1+(1-\rho)(1-\alpha)]/\{[1-(1-\rho)(1-\alpha)][1-(1-\rho)(1-\alpha)^2]\}$. This analysis shows in particular that both contributions of trapping and tumble events  yield contributions of order $\cal O(\rho)$ for $\rho\to 0$. The exact determination of all correlations $\left< \bm{a}_i\cdot\bm{a}_j\right>$ seems out of reach by the direct enumeration techniques used for $j=i+1$; we therefore focus on the small density limit $\rho\to 0$ and write
\begin{equation}
\left< \bm{a}_i\cdot\bm{a}_{i+k}\right> \underrel{\rho \to 0}{=} g(\alpha,k)\frac{\rho}{\alpha} + o(\rho),
\end{equation}
where a generalization of the argument given above for $k=1$ shows that  $\forall k,\ g(\alpha,k)\not=0$. The only two lengthscales entering this problem are $1/\rho$ and $1/\alpha$; for all $k$, $g(\alpha,k)$ has the dimension of a length squared. In the limit of low density, the only relevant lengthscale left is $1/\alpha$, and a dimensional analysis yields (taking $\alpha$ small) $g(\alpha,k)  \underrel{\alpha \to 0}{\sim} - \xi_k / \alpha^{2}$, where $\xi_k$ is a lattice-dependent dimensionless number; for $k=1$ Eq. (\ref{eq:correlations})
yields the exact value $\xi_1 = 11/24$.

Finally, the correction to the diffusion coefficient  reads
\begin{equation}
\frac{\mathcal{D} -\mathcal{D}_0}{\bar{n}/2d} \underrel{\rho,\alpha \to 0}{\sim} - \frac{2\rho}{\alpha^3}\sum_{k=1}^{\infty} \xi_k.
\label{eq:correction}
\end{equation}
 This result is exact to linear order in $\rho  $  in the limit $\alpha\to 0$. It involves  dimensionless constants $\xi_k$, which are found numerically to satisfy   $\xi_k \approx \xi_1 \Gamma^{k-1}$, with  $\Gamma\approx 0.22$. We show on Fig. \ref{fig:diffusion}(a) a perfect agreement between our numerical simulations and Eq. (\ref{eq:correction}).

We  now aim at  obtaining an approximate solution uniformly accurate in $\alpha$; to this end, we need to go beyond the linear order in $\rho/\alpha$ and therefore consider non vanishing correlations $\left< \bm{a}_i\cdot\bm{a}_{i+1}\right>$. We also generalize our argument here to mobile obstacles,  for which the nearest neighbor correlations need to be amended. Indeed,  two independent  mechanisms can now release the RTP when trapped by an obstacle, yielding correlations of a different nature: (i) a tumble of the RTP, as in the case of fixed obstacles, or (ii) a step made by the obstacle away from the course of the RTP. The latter induces large correlations in the limit $\alpha \to 0$ which must be taken into account to quantitatvely describe the RTP dynamics. Taking these events into account yields
\begin{equation}\label{gamma}
\begin{split}
\left< \bm{a}_i\cdot\bm{a}_{i+1}\right> = &\frac{\alpha}{\alpha+\rho}\frac{ \mathcal{C}_+ -\mathcal{C}_-}{2d} + \frac{\rho}{\alpha+\rho}\left(\frac{\beta^*}{\alpha^*+\beta^*} \mathcal{C}_+ \right. - \\ & \left. \frac{\alpha^*}{\alpha^*+\beta^*} \frac{\mathcal{C}_-}{2d-1} \right) \equiv \ \gamma\ell_p^2,
\end{split}
\end{equation}	
which yields the exact Eq.   (\ref{eq:correlations})  in the limit of fixed obstacles, $\beta \to 0$. In order to cover the regime of large correlations, we next assume that correlations are induced   by interactions between successive runs only;  
 classical results  \citep{hughes-book-1995} then yield $\left< \bm{a}_i\cdot\bm{a}_{j}\right> = \gamma^{|i-j|} \ell_p^2$. We have checked numerically that these correlations decay exponentially (see {\it SM}).  After summation, we obtain finally
\begin{equation}
\mathcal{D} \underrel{\rho \to 0}{\sim} \frac{ \bar n}{2d} \left[  \left< \bm{a}_i^2\right> + \frac{2\gamma}{1-\gamma}\ell_{p}^2\right]
\label{eq:fulleffecivecoeffmobile}
\end{equation}							      
where $\bar n, \left< \bm{a}_i^2\right>, \ell_p,\gamma$ are defined explicitly in Eqs. (\ref{nbar}),(\ref{a2}), and (\ref{gamma}). This explicit expression, though approximate, provides a uniformly accurate    determination  of the diffusion coefficient as we show below; it is  in addition consistent with the exact limit ($\alpha,\beta,\rho\to 0$) defined above.

%%%%%%%%
% Discussion % 
%%%%%%%%
{\it Optimized diffusivity of the RTP ---} In
Fig.~\ref{fig:diffusion}, we show our theoretical predictions for
the diffusion coefficient in the case of static and mobile
obstacles along with the results of simulations. We observe a very
good agreement between theory and simulations. For fixed obstacles
($\beta=0$), the diffusion coefficient is nonmonotonic in the
tumbling probability $\alpha$. Qualitatively, this behavior can be understood as follows:
(i) in the limit of high tumbling probability $\alpha\sim 1$, the RTP
tumbles at each time step; decreasing $\alpha $ then increases the
long time diffusion coefficient by increasing the persistence length;
(ii) in the limit of low tumbling probability $\alpha\to 0$, the RTP
proceeds mainly in long straight runs, interrupted by trapping events
whose duration $\tau_s$ diverges for $\alpha\to 0$ leading to a
vanishing diffusion coefficient. The analysis of
Eq. (\ref{eq:fulleffecivecoeffmobile}) shows that the optimal
tumbling probability satisfies $\alpha_m\propto \rho$ ; in turn, the
optimal diffusion coefficient is found to scale as
$\mathcal{D}_m\propto 1/\rho$.

In the case of mobile obstacles, the non-monotonicity in the diffusion coefficient is preserved only for a low enough obstacle mobility $\beta\le \beta_c\propto \rho$. While the diffusion coefficient in the limit of high tumbling probability is independent of the obstacle mobility in the regime $\rho \to 0$, not surprisingly, it shows a strong dependence on $\beta$ at low tumbling probability. It is interesting to note that no matter the obstacle mobility the RTP diffusion coefficient always monotonically increases for low enough decreasing tumbling probability ($\alpha \lesssim \beta$) showing that in these cases, the important unlocking mechanism is obstacle mobility.

%%%%%%%%
% Conclusion %
%%%%%%%%
{\it Discussion ---} Using a minimal model of RTPs in crowded environments, we have shown that such active particles display a nonmonotonic diffusivity as a function of the tumbling probability for static and mobile obstacles. Our analytical prediction is exact in the limit of low obstacle density for fixed obstacles. Its derivation is based on the generalization of a theorem by Kac, a strikingly simple result expected to find a variety of applications in general lattice gas problems. While derived for a particular model for which analytical progress was tractable, our results qualitatively extend far beyond this case. First, we show in {\it SM} that they extend to other microscopic types of obstacles. Then, a similar behavior has been previously observed in a mean-field model of bacterial diffusion in porous media~\citep{licata-biophysj-2016}; our result is also reminiscent of the negative differential mobility observed in Refs.~\citep{benichou-prl-2014,reichhardt-jpcm-2018} for active tracer particles. Furthermore, our results, which are exact in the low density limit, extend qualitatively to moderate to high densities of relevance to experimental settings and in particular to the diffusion of bacteria in soft agar gels~\citep{licata-biophysj-2016,croze-biophysj-2011}. Finally, while the derivation of the results reported in this Letter is specific to RTPs on lattice, the underlying mechanisms are not, and we expect similarly rich behaviors for the diffusivities of other active particles, on lattice or in continuous space. This could potentially lead to an optimization of the diffusion coefficient of active particles with respect to their reorientation dynamics and, as such, to an enhancement of their transport properties or exploration efficiency in crowded environment.

\begin{acknowledgements}
We acknowledge financial support from Institut National du Cancer (T.B. and R.V.) and from Agence Nationale de la Recherche through the grant Bacttern (Y.Z. and J.T.).
\end{acknowledgements}

%%%%%%%%%%%%%%
%\bibliographystyle{apsrev4-1}
%\bibliography{biblio}
%%%%%%%%%%%%%%


%%%%%%%%%% Merge with supplemental materials %%%%%%%%%%
\onecolumngrid
\pagebreak
\begin{center}
\textbf{\large Supplemental Material for "Optimized Diffusion of Run-and-Tumble Particles in Crowded Environments"}
\end{center}
%%%%%%%%%% Merge with supplemental materials %%%%%%%%%%
%%%%%%%%%% Prefix a "S" to all equations, figures, tables and reset the counter %%%%%%%%%%
\setcounter{equation}{0}
\setcounter{figure}{0}
\setcounter{table}{0}
\setcounter{page}{1}
\makeatletter
\renewcommand{\theequation}{S\arabic{equation}}
\renewcommand{\thefigure}{S\arabic{figure}}
\renewcommand{\bibnumfmt}[1]{[S#1]}
\renewcommand{\citenumfont}[1]{#1}
%%%%%%%%%% Prefix a "S" to all equations, figures, tables and reset the counter %%%%%%%%%%

%%%%%%%%%%%%%%%%%%%%%
% Schematic description of an obstacle  %
%%%%%%%%%%%%%%%%%%%%%
\section{Schematic description of an obstacle}

\begin{figure}[h]
\centering
\includegraphics[width=0.4\textwidth]{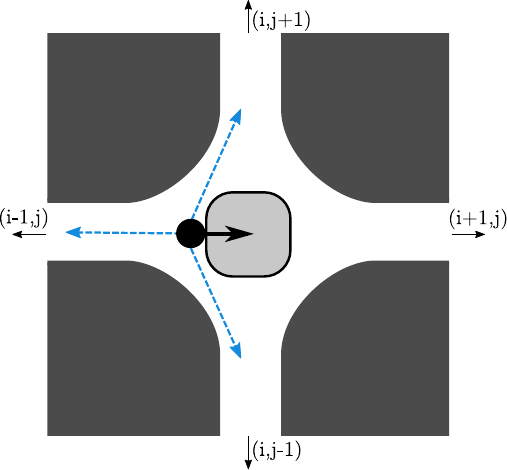}
\caption{{\it Schematic description of the obstacles in our lattice gas model}}
\label{fig:obstacle_1}
\end{figure}

In this section, we illustrate on Figure \ref{fig:obstacle_1} the definition of the obstacles we consider in the main text. We consider obstacles (in light grey) interacting via hard-core repulsion, {\it i.e.} each lattice site can contain at most one obstacle. The run-and-tumble particle (RTP) is assumed to have hard-core interactions with obstacles; we consider its size to be negligible in front of the size of the obstacles. The RTP can therefore jump on a lattice site occupied by an obstacle. When a RTP (black dot with polarity along the black arrow) arrives on a site (i,j) occupied by an obstacle, the latter prevents the RTP from going onwards along its polarity. The RTP only resumes its motion after either: (i) a tumble leading to one of the escape routes highlighted by the light blue dashed arrows or (ii) if the obstacles diffuses to another lattice site.

%%%%%%%%%%%%%%%%%%%%%%%%%%%%%%%%%%
% Generalization of Wald's identity and Equation (2) of main text   %
%%%%%%%%%%%%%%%%%%%%%%%%%%%%%%%%%%
\section{Generalization of Wald's identity and Equation (2) of main text}

We start from the decomposition of the trajectories of the RTP in a sequence of linear runs $\bm{a}_i$, punctuated by either encounters with obstacles, or tumbles in any direction, as defined in the main text:
\begin{equation}
\bm{r}(t) = \sum_{i=1}^{n(t)} \bm{a}_i.
\label{Seq:decomposition1}
\end{equation}
Here the random variable $n(t)$ denotes the number of linear runs composing the trajectory up to a given time $t$. We introduce the probability $p_t(n)$ that $n$ linear runs occurred up to time $t$. 
We then write 
\begin{equation}
\left< r^2(t) \right> =\sum_{n=0}^{+\infty}\left< r^2(t) \right>_n p_t(n)
\label{Seq:decomposition2}
\end{equation}
where $\left< r^2(t) \right>_n$ denotes the average conditioned by $n$. It follows that
\begin{equation}
\left< r^2(t) \right> =\left< n(t) \right>\left< \bm{a}_i^2 \right>+\sum_{n=0}^{+\infty} p_t(n)\sum_{\substack{i,j=1 \\ i \ne j} }^{n} \left< \bm{a}_i\cdot\bm{a}_j\right>.
\label{Seq:decomposition2}
\end{equation}
Then one has
\begin{equation}
\left|\sum_{n=0}^{+\infty} p_t(n)\sum_{\substack{i,j=1 \\ i \ne j} }^{n} \left< \bm{a}_i\cdot\bm{a}_j\right>-\sum_{\substack{i,j=1 \\ i \ne j} }^{\left< n\right>} \left< \bm{a}_i\cdot\bm{a}_j\right>\right|\le \left< \bm{a}_i^2 \right>\left< |n-\left< n\right>| \right>\underrel{t \to \infty}{\ll} \left< n(t) \right>\left< \bm{a}_i^2 \right>.
\label{Seq:decomposition2}
\end{equation}
Here we have used the fact that $\left< n(t) \right>\underrel{t \to \infty}{\sim} {\bar n} t$, as justified in the main text, and $\sqrt{\left< (n(t)-\left< n(t) \right>)^2 \right>}/\left< n(t) \right>\underrel{t \to \infty}{\to} 0$. Finally, this yields Equation (2) of the main text:

\begin{equation}
\left< r^2(t) \right> \underrel{t \to \infty}{\sim} \left< n(t) \right>\left< \bm{a}_i^2 \right> + \sum_{\substack{i,j=1 \\ i \ne j} }^{\left< n(t) \right>} \left< \bm{a}_i\cdot\bm{a}_j\right> .
\label{Seq:decomposition2}
\end{equation}

%%%%%%%%%%%%%%%%%%%%%%%%%%%%%%%%%%%%%%%%
% Application of Kac's theorem on the mean return time of Markov processes  %
%%%%%%%%%%%%%%%%%%%%%%%%%%%%%%%%%%%%%%%%
\section{Application of Kac's theorem on the mean return time of Markov processes}

The general statement of Kac theorem for the mean return time of Markov chain can be found in \citep{aldous-mono-2002}. This theorem takes a simple form in the case of irreducible finite state discrete time Markov processes $(X_t)$, if one considers a subset $A$ of states of interest such that $X_t\in A \to X_{t+1}\not\in A$ : the process $X_t$ spends a single time step in $A$ at each visit of $A$. The Kac theorem then gives that the mean of the  return time $T_r$ defined by $T_r={\rm inf}\left\{t\ge 1:X_0\in A, X_t\in A\right\}$ is exactly given by
\begin{equation}
\langle T_t\rangle=1/P_s(A)
\end{equation}
where $P_s(A)$ is the stationary measure of the set $A$. In the case of the RT dynamics with obstacles that we study in this paper, in order to apply Kac theorem in this simple form, we had to  introduce the auxiliary process  $\bm{\tilde r}(t)$. It  is identical to  $\bm{r}(t)$ in mobile phases, but the durations of all its static phases are set to 1 : upon each trapping event by  an obstacle the auxiliary process is released in the same direction as the original process $\bm{ r}(t)$ would be, but  after a single time step. The mean running time is then identical for both the original process of interest   $\bm{r}(t)$ and $\bm{\tilde r}(t)$; for the process $\bm{\tilde r}(t)$, which has a uniform stationary distribution, Kac theorem is directly applicable and yields $ \left< \tau_r \right>=1/P_{\rm stat}({\cal O})$, where $P_{\rm stat}({\cal O})$ is the stationary probability of $\cal O$ for the auxiliary process.  We therefore find the simple expression
\begin{equation}
\left< \tau_r \right> = \frac{1}{P_{\mathrm{stat}}({\cal O})} = \frac{1}{\rho},
\label{Seq:kac}
\end{equation}
which is strikingly independent of the tumbling probability $\alpha$. It should be noted that the position process $\bm{r}(t)$ is not strictly Markovian, because   the polarity of the particle needs to be specified to obtain a Markovian evolution. The process $\bm{r}(t)$ being nevertheless ergodic Kac theorem still applies.

\begin{figure}[h]
\centering
\includegraphics[width=0.46\textwidth]{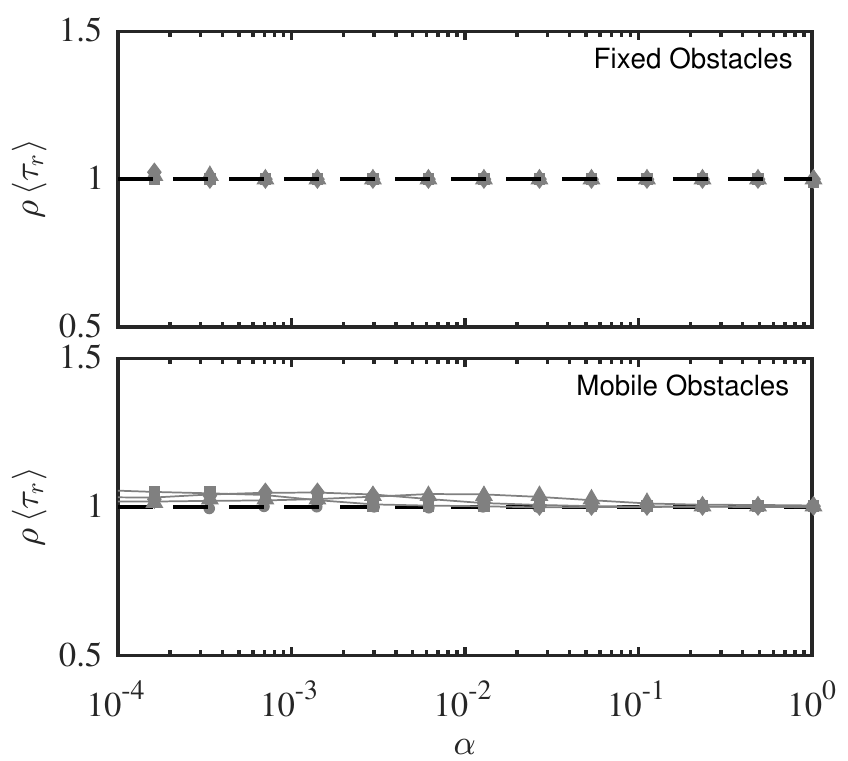}
\caption{{\it Running times} --- Average running time $\left< \tau_R \right>$ on a semilog scale renormalized by Kac's prediction $1/\rho$ for: ({\it Top}) Fixed obstacles ($\beta=0$) at various densities $\rho = 10^{-3}$ ({\it squares}), $10^{-2}$ ({\it diamonds}) and $10^{-1}$ ({\it triangles}); ({\it Bottom}) Mobile obstacles at fixed density $\rho = 10^{-2}$ for various mobilities $\beta = 0$ ({\it circles}), $10^{-4}$ ({\it squares}), $10^{-3}$ ({\it diamonds}) and $10^{-2}$ ({\it triangles}). The dashed lines are the theoretical predictions.}
\label{fig:timing}
\end{figure}

%%%%%%%%%%%%%%%%%%%%%%%%%%
% Derivation of the nearest neighbor correlations  %
%%%%%%%%%%%%%%%%%%%%%%%%%%
\section{Derivation of the nearest neighbor correlations}
In this section, we provide additional details about the derivation of the nearest neighbor correlations $\left< \bm{a}_i\cdot\bm{a}_{i+1}\right>$. We only consider here the case of fixed obstacles. In particular, in steady state, every site has the same occupation probability $\rho$ and the two mechanisms terminating a run (tumble or encounter with an obstacle) are statistically independent (this problem can be seen as an infinite sequence of Bernoulli trials); we can write the effective persistence length as follows
\begin{equation}
\ell_p \equiv \left< | \bm{a}_i | \right> = \sum_{n=1}^{\infty} n(1-\rho)^{n-1}(1-\alpha)^{n-1}[1-(1-\rho)(1-\alpha)] = \frac{1}{1-(1-\rho)(1-\alpha)}
\end{equation}

As in the main text, we can partition these correlations between the two types of events ending straight run $i$: (1) a {\it tumble} and (2) an {\it encounter with an obstacle} followed by an unlocking tumble. Thus, we write
\begin{equation}
\left< \bm{a}_i\cdot\bm{a}_{i+1}\right> = w_t \left< \bm{a}_i\cdot\bm{a}_{i+1}\right>_t + w_o \left< \bm{a}_i\cdot\bm{a}_{i+1}\right>_o
\label{Seq:corr_tot}
\end{equation}

with $w_t$ and $w_o$ the probabilities that the run ended in a tumble and an encounter with an obstacle respectively. By definition, these probabilities read
\begin{align}
w_t &\equiv \frac{\alpha}{\alpha+\rho} \\
w_o &\equiv \frac{\rho}{\alpha+\rho}
\end{align}

On one hand, a straight run ending in an encounter with an obstacle (followed by an unlocking tumble) can only lead to negative correlations, as the following run needs to be polarized in the exact opposite direction to have a non-zero contribution. Indeed, orthogonal directions would lead to null contributions and a tumble ending in the TP polarized in the original direction would not be unlocking (only $2d-1$ directions are unlocking the TP). Hence, we can write
\begin{equation}
\left< \bm{a}_i\cdot\bm{a}_{i+1}\right>_o = \frac{-\mathcal{C}_-}{2d-1} 
\label{Seq:corr_o}
\end{equation}

On the other hand, a straight run ending in a tumble (with no obstacle) will lead to both positive and negative contributions. Indeed, a tumble can lead to a non-zero contribution by polarizing the TP either in the original polarization (positive correlations) or in the opposite direction to the original polarization (negative correlations) with same probability $1/(2d)$, any other direction of the new polarization will lead to zero contribution to the correlations. As such, we write
\begin{equation}
\left< \bm{a}_i\cdot\bm{a}_{i+1}\right>_t = \frac{\mathcal{C}_+ - \mathcal{C}_-}{2d} 
\label{Seq:corr_t}
\end{equation}

First, we consider the positive correlations between runs $i$ and $i+1$. The fundamental process we are modelling here is Markovian, as such two following runs are independent and as a consequence, we can write
\begin{equation}
\mathcal{C}_+ = \left< |\bm{a}_i|\right>\left< |\bm{a}_{i+1}|\right> = \left< |\bm{a}_i|\right>^2 = \ell_p^2 = \frac{1}{[1-(1-\rho)(1-\alpha)]^2}
\end{equation}

Let us now turn to the negative correlations which require a bit more work. Indeed, negative correlations come from runs in opposite directions. Qualitatively, for all steps of the run $i+1$ overlapping with run $i$, the TP will not encounter any obstacle with probability $1$ and are therefore favored. Here, we partition the correlations as a function of the event starting the run $i$ and we write
\begin{equation}
\begin{split}
\mathcal{C}_- = &\sum_{n=1}^{\infty} n(1-\rho)^{n-1}(1-\alpha)^{n-1}\rho \left[ \sum_{m=1}^{n-1}m(1-\alpha)^{m-1} \alpha + n(1-\alpha)^{n-1} \right] + \\
			&\sum_{n=1}^{\infty} n(1-\rho)^{n-1}(1-\alpha)^{n-1}\alpha(1-\rho) \left[ \sum_{m=1}^{n}m(1-\alpha)^{m-1} \alpha \right.  \\ &+ \left. \sum_{m=n+1}^{\infty} m(1-\alpha)^{m-1}(1-\rho)^{m-n-1} [1-(1-\rho)(1-\alpha)] \right]
\end{split}
\label{Seq:negcorr}
\end{equation}

The first term in Equation \ref{Seq:negcorr} corresponds to the runs $i$ starting by unlocking from an obstacle. In this case, we know that the run $i+1$ can be at most as long as run $i$. We also know that, for the whole length of run $i+1$, the TP will not encounter any obstacle and the only possible event is a tumble of the RTP. The second term corresponds to trajectories with run $i$ starting with a tumble (and no obstacle). Here, when the run $i$ is of length $n$, we know with certainty that the TP will not encounter any obstacle for its first $n$ steps in run $i+1$. In this case, run $i+1$ is not limited to be at most as long as run $i$ and can thus exceed the length of run $i$; as a consequence, for runs $i$ of length $n$, we partition the correlations over runs $i+1$ of length $m \leq n$ and runs $i+1$ of length $m \ge n+1$ and give them proper weights. One can easily check that these conditional probabilities are properly normalized. Finally, Equation \ref{Seq:negcorr} reduces to
\begin{equation}
\mathcal{C}_- = \frac{1+(1-\rho)(1-\alpha)}{[1-(1-\rho)(1-\alpha)][1-(1-\rho)(1-\alpha)^2]}
\end{equation}

\begin{figure*}[t]
\centering
\includegraphics[width=0.95\textwidth]{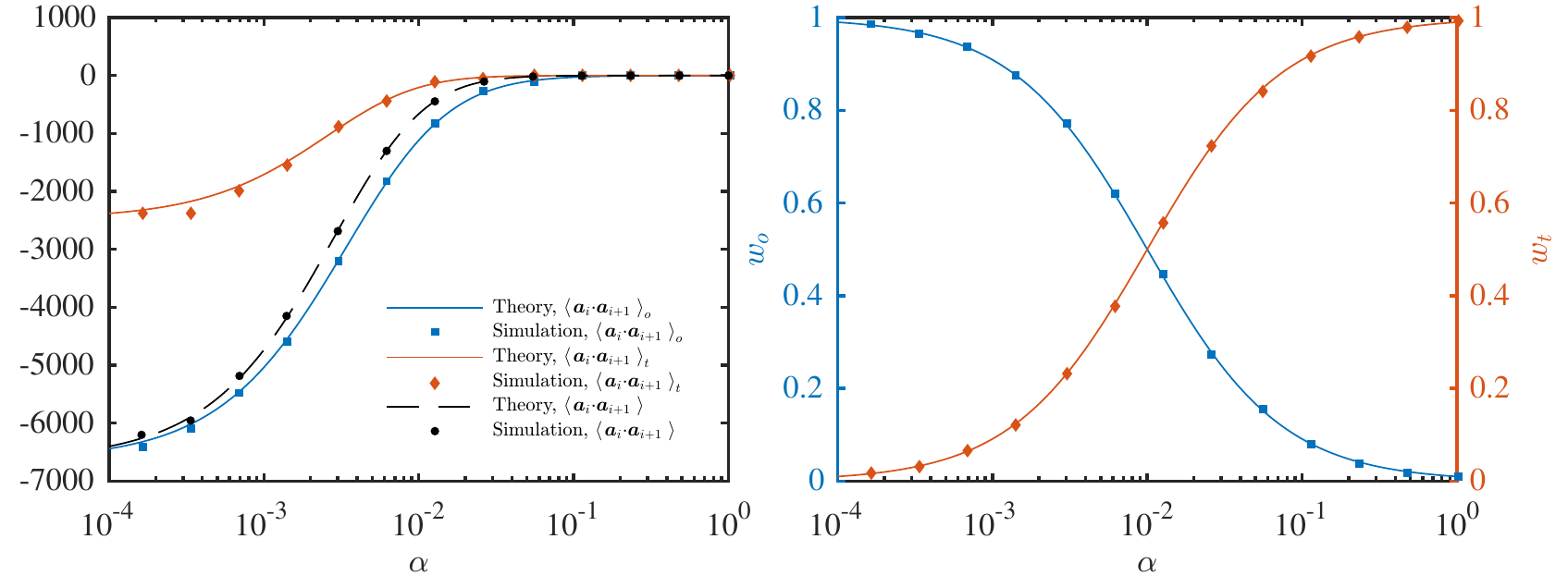}
\caption{{\it Nearest neighbor correlations} ---  ({\it Left}) Decomposition of all terms in correlations between runs $i$ and $i+1$ from Equations \ref{Seq:corr_o}, \ref{Seq:corr_t} and \ref{Seq:corr_tot} (lines) and from numerical simulations (symbols - averaged over long runs for at least 1000 different obstacle configurations) displayed as a function of the tumbling probability $\alpha$; ({\it Right}) Weight of each type of correlations (coming from a simple tumble or an encounter with an obstacle) from their definition (solid lines) and as measured in numerical simulations (symbols). All data and theoretical predictions obtained for $\rho = 0.01$.}
\label{fig:nearcorr}
\end{figure*}

Figure \ref{fig:nearcorr} shows a perfect agreement between the nearest neighbor correlation terms as calculated from our theoretical prediction and measured in numerical simulations. Further, we find that 
\begin{equation}
\lim_{\rho \to 0} \mathcal{C}_- = \frac{1}{\alpha^2}
\end{equation}
which is consistent with the fact that when $\rho \to 0$, the successive runs are all independent of each other and thus $\left< \bm{a}_i\cdot\bm{a}_{i+1}\right> \underrel{\rho \to 0}{\to} 0$. In the limit of low tumbling probability (for finite density), we find unsurprisingly that only the negative correlations term matters and we have
\begin{equation}
\lim_{\alpha \to 0} \left< \bm{a}_i\cdot\bm{a}_{i+1}\right> = \frac{1}{2d-1}\frac{\rho-2}{\rho^2}
\end{equation}

To go one step further, we expand the nearest neighbor correlations in power series of the density $\rho$ and we obtain
\begin{equation}
\left< \bm{a}_i\cdot\bm{a}_{i+1}\right> \underrel{\rho \to 0}{=} g(\alpha,1)\frac{\rho}{\alpha} + o(\rho)
\end{equation}

with 
\begin{equation}
g(\alpha,1) = \frac{3\alpha^2 -10 \alpha +11}{12\alpha^2(\alpha-2)}
\end{equation}

Finally, we find by taking first the limit of low density ($\rho \to 0$) and next the limit of low tumbling probability ($\alpha \to 0$), always keeping $\rho \ll \alpha$, the correlations between successive runs read
\begin{equation}
\left< \bm{a}_i\cdot\bm{a}_{i+1}\right> \underrel{\rho,\alpha \to 0}{=} -\frac{\xi_1}{\alpha^2}\frac{\rho}{\alpha} + o(\rho)
\end{equation}

with $\xi_1 = 11/24$. This result is consistent with the fact that $\rho/\alpha$ is the only non-dimensional parameter in this problem and that in the limit $\rho \to 0$ (free diffusion), the only lengthscale in the system is $1/\alpha$. By dimensional analysis, one obtains that the correlations between successive runs must scale as $\rho / \alpha^3$.

%%%%%%%%%%%%%%%%%%%%%%%%%%%%%%
% Derivation of the diffusion coefficient for fixed obstacles %
%%%%%%%%%%%%%%%%%%%%%%%%%%%%%%
\section{Derivation of the diffusion coefficient for fixed obstacles}

The above argument can be generalized to higher order correlations, even though an explicit calculation seems out of reach. We find that higher order correlations are also linear in the lowest order of the obstacle density
\begin{equation}
\left< \bm{a}_i\cdot\bm{a}_{i+k}\right> \underrel{\rho \to 0}{=} g(\alpha,k)\frac{\rho}{\alpha} + o(\rho)
\end{equation}

In the limit of low tumbling probabilities (with $\rho \ll \alpha$), we found that $g(\alpha,k)  \underrel{\alpha \to 0}{\sim} - \xi_k / \alpha^{2}$. The diffusion coefficient is given by
\begin{equation}
\mathcal{D} \equiv \lim_{t \to \infty} \frac{\left< r^2(t)\right>}{2dt}
\end{equation}
with $\left< r^2(t) \right> \underrel{t \to \infty}{\sim} \left< n(t) \right>\left< \bm{a}_i^2 \right> + \sum_{i,j=1}^{\left< n(t) \right>} \left< \bm{a}_i\cdot\bm{a}_j\right>$. In the long time limit, we can rewrite this
\begin{equation}
\left< r^2(t) \right> \underrel{t \to \infty}{\sim} \left< n(t) \right> \left[ \left< \bm{a}_i^2 \right> + 2\sum_{k=1}^{\infty} \left< \bm{a}_i\cdot\bm{a}_{i+k}\right> \right].
\end{equation}

We can sum over all possible correlations and the diffusion coefficient thus reads
\begin{equation}
\mathcal{D} \underrel{\rho,\alpha \to 0}{\sim} \frac{\bar{n}}{2d} \left[ \left< \bm{a}_i^2\right> -\frac{2}{\alpha^2} \frac{\rho}{\alpha}\sum_{k=1}^{\infty} \xi_k \right]
\end{equation}

with $\left< a_i^2\right> = [1+(1-\rho)(1-\alpha)]/[1-(1-\rho)(1-\alpha)]^2$. This result is exact in the limit $\rho \to 0$ for static obstacles. In practice, we find a good approximation in $\xi_k = \xi_1 \Gamma^{k-1}$ when $\alpha \to 0$. Finally, in the limit of $\rho \to 0$, $\alpha \to 0$ with $\rho<<\alpha$, the diffusion coefficient reads
\begin{equation}
\mathcal{D} \approx \frac{\bar{n}}{2d}  \left[ \left< \bm{a}_i^2\right> - \frac{2\rho}{\alpha^3}\frac{\xi_1}{1-\Gamma}\right]
\label{Seq:fulleffecivecoeff}
\end{equation}
where we determine numerically $\Gamma \approx 0.22$.

%%%%%%%%%%%%%%%%%%%%%%%%%%%%%%%
% Derivation of the diffusion coefficient for mobile obstacles %
%%%%%%%%%%%%%%%%%%%%%%%%%%%%%%%
\section{Derivation of the diffusion coefficient for mobile obstacles}

In the main text, we generalize our argument to mobile obstacles for which the nearest neighbor correlations need to be amended. In this case, we assume that the obstacles perform nearest neighbour random walks, with probability $\beta<1$ to jump at each time step. Indeed, the two independent unlocking mechanisms to escape an obstacle will lead to correlations between successive runs, in particular: (1) if the RTP tumbles and ends up with a polarity in the opposite direction from the obstacle leading to negative correlations, (2) the obstacle moves in one of the directions perpendicular with the polarity of the TP leading to positive correlations. We neglect here the correlations between successive runs if the obstacle moves in the direction of the RTP polarity; indeed, in this case, the RTP would immediately be trapped at the next step leading to correlations of the order $\ell_p$, negligible in front of the other correlation terms  of the order of $\ell_p^2$. Thus, we can partition over both termination mechanisms and unlocking mechanisms and we obtain
\begin{equation}
\left< \bm{a}_i\cdot\bm{a}_{i+1}\right> = \frac{\alpha}{\alpha+\rho}\frac{ \mathcal{C}_+ -\mathcal{C}_-}{2d} + \frac{\rho}{\alpha+\rho}\left[\frac{\beta^*}{\alpha^*+\beta^*} \mathcal{C}_+ -\frac{\alpha^*}{\alpha^*+\beta^*} \frac{\mathcal{C}_-}{2d-1} \right] \equiv \ \gamma \ell_p^2
\label{Seq:corr_beta}
\end{equation}	
with probabilities $\alpha^* = \alpha(2d-1)/2d$ and $\beta^* = \beta(2d-1)/2d$. The first term in Equation \ref{Seq:corr_beta} remains unchanged, while the second term now contains the two possible unlocking mechanisms when the RTP encounters an obstacle. If we consider the theory of persistent walks, we know that more generally the correlations decay exponentially with the distance between runs and are given by
\begin{equation}
 \left< \bm{a}_i\cdot\bm{a}_{i+k}\right>  = \ell_{p}^2 \gamma^{|k|}
\end{equation}

\begin{figure*}[h]
\centering
\includegraphics[width=0.46\textwidth]{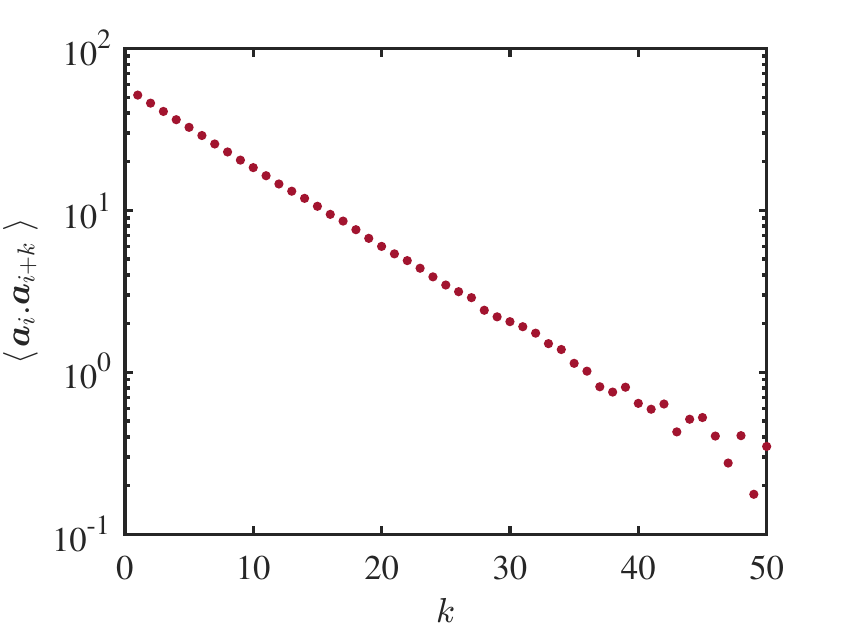}
\caption{{\it Correlations between straight runs} --- exponential decay of correlations between straight runs averaged over 1000 realizations of simulations at obstacle density $\rho = 0.1$, tumbling probability $\alpha = 0.001$ and obstacle mobility $\beta = 0.01$.}
\label{fig:decay}
\end{figure*}

For instance, Figure \ref{fig:decay} shows a clear exponential decay of these correlations in the case of a simulation with obstacle density $\rho = 0.1$, tumbling probability $\alpha = 0.001$ and obstacle mobility $\beta = 0.01$. Thus, in the long time limit, we obtain
\begin{align}
\sum_{i \ne j}  \left< \bm{a}_i\cdot\bm{a}_{j}\right> &= \ell_{p}^2  \sum_{i \ne j} \gamma^{|i-j|} \underrel{t \to \infty}{\sim}  \left<n(t)\right> \frac{2\gamma}{1-\gamma} \ell_{p}^2
\end{align}

Finally, in this case, the diffusion coefficient reads: 
\begin{equation}
\mathcal{D} \equiv \lim_{t \to \infty} \frac{\left< r^2(t) \right>}{2dt} = \frac{\bar{n}}{2d} \left[ \left< \bm{a}_i^2\right> +\frac{2\gamma}{1-\gamma}\ell_{p}^2\right]
\end{equation}

%%%%%%%%%%%%%%
% Numerical Simulations %
%%%%%%%%%%%%%%
\section{Numerical Simulations}

We perform Monte-Carlo simulations on a two-dimensional square lattice of step size 1. At $t=0$, the RTP is placed at the origin and we seed the lattice randomly (excluding the origin) such that the density of obstacles is uniform over the domain. For a given density $\rho$, we know that the total number of obstacles $N_{\mathrm{obs}}$ is given by $N_{\mathrm{obs}} = \rho L^2$ with $L$ the width of the lattice. To avoid finite size effect when $\alpha \to 0$, we ensure that our configurations have on average 10 obstacles per row. This imposes that $N_{\mathrm{obs}} / L = \rho L \approx 10$. Thus, this condition dictates the total size of the lattice $L^2$ which we have checked to be large enough to be considered infinite for our purposes (for instance, for $\rho = 0.01$ corresponds to a lattice of size $1000\times1000$). At each simulation time step, we start by checking for a tumble event and then we simultaneously evolve the position of the obstacles and the TP. Running times $\tau_r$ are defined from the moment the TP leave an obstacle, until the moment they step on the next obstacle; waiting times $\tau_s$ are defined from the moment the TP steps on an obstacle to the moment when it releases itself. For all data presented in the main text, statistics are obtained over at least 1000 realizations and calculated over trajectories of $10^6$ steps. As a consequence, the simulations were run on timescales much larger than any relevant relaxation timescale, placing our simulations in a stationary state. This process can be assumed to be ergodic, and ensemble and time averages to be equivalent.

%%%%%%%%%%%%%%%%%%%%%%%%%%%%%%%%%%%%%%%%%%%%%%%%%%%%%%%%%%%%
% Diffusivity of a Run-and-tumble particle interacting with the obstacles via hard-core repulsion with excluded volume   %
%%%%%%%%%%%%%%%%%%%%%%%%%%%%%%%%%%%%%%%%%%%%%%%%%%%%%%%%%%%%
\section{Diffusivity of a Run-and-tumble particle interacting with the obstacles via hard-core repulsion with excluded volume}

\begin{figure}[h]
\centering
\includegraphics[width=0.55\textwidth]{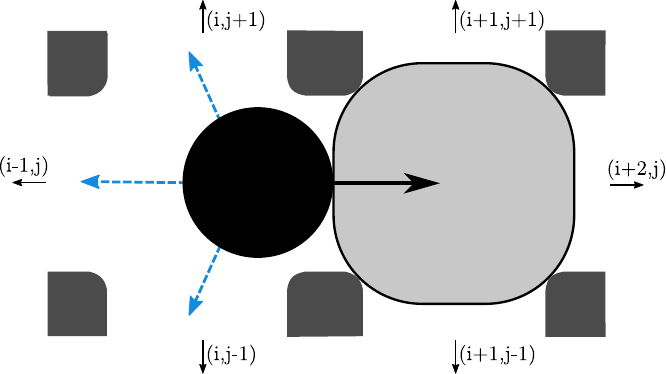}
\caption{{\it Schematic description of the obstacles in the case where the RTP is not small compared to the obstacle} ---  The RTP cannot enter the lattice site occupied by an obstacle. As for Figure \ref{fig:obstacle_1}, the obstacle is depicted in light grey, the RTP as a black disk with polarity along the black arrow and the three escape routes following a tumble of the RTP are pictured here by the dashed blue lines.}
\label{fig:obstacle_2}
\end{figure}

In the main text, we made the assumption that the size of the RTP was small compared to the size of the obstacles, leading to a trapping mechanism where the RTP can jump on a lattice site already occupied by an obstacle, but gets stuck there (see Figure \ref{fig:obstacle_1} in {\it Supplemental Material}). To check the genericity of our results, we performed simulations in the case where the RTP cannot even enter a site occupied by an obstacle as shown on Figure \ref{fig:obstacle_2}. 

\begin{figure*}[h]
\centering
\includegraphics[width=0.46\textwidth]{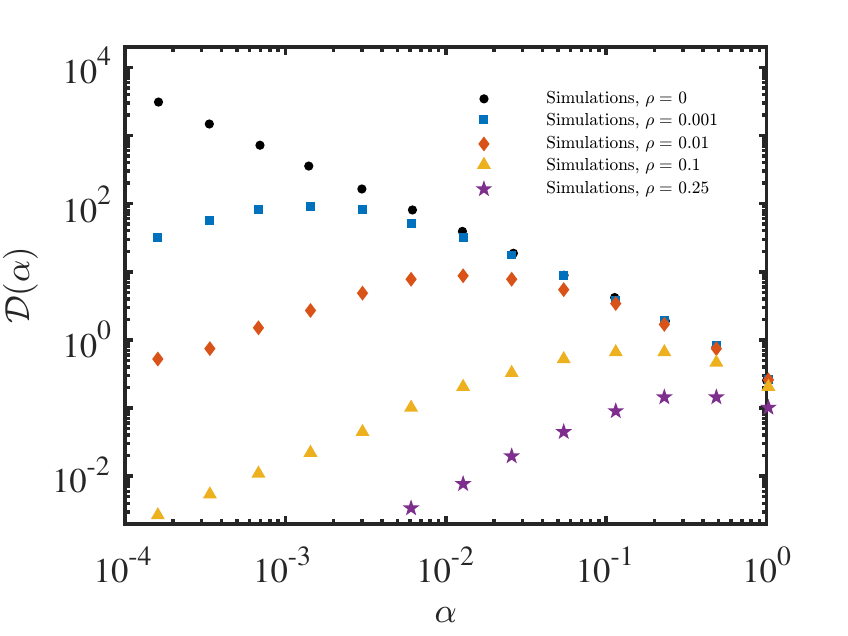}
\caption{{\it Diffusivity of a run-and-tumble particle} --- interacting with fixed obstacles via hard-core repulsion at various obstacle densities $\rho$ as a function of tumbling probability $\alpha$. These results are qualitatively similar to results presenting in Figure 2 of the main text.}
\label{fig:diffhc}
\end{figure*}

In this case, the RTP does not get trapped on the same lattice site as the trapping obstacle but rather on the neighboring lattice site. We show on Figure \ref{fig:diffhc} the diffusivity of the RTP. These results are qualitatively similar to those presented in the main text; the diffusivity shows a non-monotonic behavior and displays an optimum as a function of the tumbling probability $\alpha$. The tumbling probability for optimal diffusivity depends on the density of obstacles $\rho$ in the same manner as shown in the main text, {\it i.e.} the optimal tumbling probability increases with the density of obstacles. It is important to note that, as shown on Figure \ref{fig:diffhc}, the existence of this optimal tumbling probability persists at moderate to high densities (tested numerically up to $\rho = 0.25$).

\end{document}